\documentclass[conference]{IEEEtran}
\IEEEoverridecommandlockouts
\usepackage[noadjust]{cite}
\usepackage{amsmath,amssymb,amsfonts,amsbsy}
\usepackage{algorithmic}
\usepackage{graphicx}
\usepackage{textcomp}
\usepackage{xcolor}
\usepackage{enumitem}
\usepackage{etoolbox}
\usepackage{multirow}
\usepackage{svg}
\usepackage{arydshln}
\usepackage{mathtools}
\usepackage{subfig}
\usepackage{caption}
\usepackage{subcaption}
\usepackage{amsmath}
\usepackage{cite}

\usepackage{caption,subcaption}
\usepackage{caption}
\usepackage{subcaption}
\usepackage{dblfloatfix}
\usepackage{amsmath,amssymb,amsfonts,amsbsy,amsthm}

\usepackage{array}

\newcolumntype{C}[1]{>{\centering\let\newline\\\arraybackslash\hspace{0pt}}m{#1}}

\usepackage{hyperref}
\usepackage{array}
\newcolumntype{P}[1]{>{\centering\arraybackslash}p{#1}}
\makeatletter
\newcommand*{\rom}[1]{\expandafter\@slowromancap\romannumeral #1@}
\makeatother
\usepackage{amsthm}
\theoremstyle{definition}

\usepackage{dblfloatfix}
\usepackage{tabularx,booktabs}
\newcolumntype{Y}{>{\centering\arraybackslash}X}
\newcolumntype{b}{X}
\newcolumntype{s}{>{\hsize=.75\hsize}X}
\newcolumntype{t}{>{\hsize=1.25\hsize}X}
\def\BibTeX{{\rm B\kern-.05em{\sc i\kern-.025em b}\kern-.08em
    T\kern-.1667em\lower.7ex\hbox{E}\kern-.125emX}}

\usepackage{graphicx,wrapfig}

\usepackage{array}
\newcolumntype{L}{>{\centering\arraybackslash}m{3cm}}
\usepackage{url}
\makeatletter
\g@addto@macro{\UrlBreaks}{\UrlOrds}
\makeatother

\usepackage[singlespacing]{setspace} 
\usepackage{listings}
\usepackage{xcolor}

\lstdefinelanguage{json}{
    morestring=[b]",
    morekeywords={true,false,null},
    sensitive=true,
    morecomment=[l]{//}
}

\usepackage{bbding}
\usepackage{fontawesome5}

\begin{document}


\title{
Empirical Study of Post-Quantum Cryptography Adoption Status in SSH Infrastructure 
}

\author{\IEEEauthorblockN{\textsuperscript{} Md Ariful Haque\textsuperscript{(\faEnvelope[regular])}, Sander Vinberg, Ben Edwards, Jo\~ao Cruz}
\IEEEauthorblockA{\textit{Bitsight Technologies, Boston, MA, USA} \\
\{mdariful.haque, sander.vinberg, benjamin.edwards, joao.cruz\}@bitsighttech.com}
}

\maketitle

\makeatletter
\newif\if@borderstar
\def\bordermatrix{\@ifnextchar*{%
\@borderstartrue\@bordermatrix@i}{\@borderstarfalse\@bordermatrix@i*}%
}
\def\@bordermatrix@i*{\@ifnextchar[{\@bordermatrix@ii}{\@bordermatrix@ii[()]}}
\def\@bordermatrix@ii[#1]#2{%
\begingroup
\m@th\@tempdima8.75\p@\setbox\z@\vbox{%
\def\cr{\crcr\noalign{\kern 2\p@\global\let\cr\endline }}%
\ialign {$##$\hfil\kern 2\p@\kern\@tempdima & \thinspace %
\hfil $##$\hfil && \quad\hfil $##$\hfil\crcr\omit\strut %
\hfil\crcr\noalign{\kern -\baselineskip}#2\crcr\omit %
\strut\cr}}%
\setbox\tw@\vbox{\unvcopy\z@\global\setbox\@ne\lastbox}%
\setbox\tw@\hbox{\unhbox\@ne\unskip\global\setbox\@ne\lastbox}%
\setbox\tw@\hbox{%
$\kern\wd\@ne\kern -\@tempdima\left\@firstoftwo#1%
\if@borderstar\kern2pt\else\kern -\wd\@ne\fi%
\global\setbox\@ne\vbox{\box\@ne\if@borderstar\else\kern 2\p@\fi}%
\vcenter{\if@borderstar\else\kern -\ht\@ne\fi%
\unvbox\z@\kern-\if@borderstar2\fi\baselineskip}%
\if@borderstar\kern-2\@tempdima\kern2\p@\else\,\fi\right\@secondoftwo#1 $%
}\null \;\vbox{\kern\ht\@ne\box\tw@}%
\endgroup
}
\makeatother

\begin{abstract}
With the anticipated arrival of Q-day, quantum computers threaten current cryptographic schemes for encryption and digital signatures. In response, major content delivery networks and cloud service providers are already adopting Post-Quantum Cryptography (PQC). The current state-of-the-art focuses on benchmarking the operational performance of implementing PQC, but the real-world adoption status remains under-researched. This work utilizes large-scale SSH fingerprint data from a proprietary scanning tool to analyze PQC adoption across 1.52 million hosts and 27,012 entities. Our study shows that the PQC transition in the SSH protocol is underway but highly polarized: 45.2\% of SSH hosts on port 22 support PQC, while 54.8\% remain classical-only. At the entity level, 39.9\% have enabled PQC on at least one host, whereas 60.1\% have no PQC support. By correlating PQC adoption with industry, geography, and organizational size (digital and financial), this study provides a comprehensive empirical assessment of PQC adoption status in SSH infrastructure.
\end{abstract}

\begin{IEEEkeywords}
Post-Quantum Cryptography, Cybersecurity, Encryption Algorithms, Secure Shell, Empirical Analysis, ML-KEM, Streamlined NTRU Prime, Internet Scanning.
\end{IEEEkeywords}

%
\section{Introduction}
Today, communication and information exchange between parties over the internet are secured using cryptographic techniques, including encryption and digital signatures, ensuring confidentiality and non-repudiation. Cryptographic algorithms have long been considered secure, as the underlying mathematical formulations of popular Diffie-Hellman, Elliptic-Curve cryptography, or RSA algorithms are based on large prime factorization, which is computationally infeasible to reverse using traditional computing resources. However, the emergence of quantum computers is expected to drastically reduce the time required to break such encryption from years to days. 

Quantum computers use quantum mechanics, where the fundamental unit of information is a Qubit (quantum bit, aka Q-bit), which can represent 0, 1, or a `superposition' of both states simultaneously. Unlike classical computers (binary switches), qubits use quantum mechanics, where properties like `quantum entanglement' can perform complex calculations exponentially faster. Shor’s algorithm \cite{shor1999polynomial} can factor large prime numbers exponentially faster than classical computers. Also, Grover’s algorithm \cite{grover1996fast} can significantly weaken the security strength of symmetric encryption techniques (e.g., AES-128). There is also the risk of the ``harvest-now, decrypt-later (HNDL)'' threat, a surveillance strategy that relies on the acquisition and long-term storage of currently unreadable encrypted data awaiting possible breakthroughs in decryption technology. For data requiring long-term confidentiality, this poses a serious threat. 

With a multi-year effort at the national and global level, the National Institute of Standards and Technology (NIST) has finally announced three post-quantum cryptographic standards in August 2024 \cite{nist2024pqc}; one for general encryption - FIPS (Federal Information Processing Standard) 203, aka CRYSTALS-Kyber (later renamed as ML-KEM, i.e., Module-Lattice-Based Key-Encapsulation Mechanism); the other two are for protecting digital signatures: FIPS 204 (aka CRYSTALS-Dilithium, later renamed as ML-DSA i.e., Module-Lattice-Based Digital Signature Algorithm) and FIPS 205 (based on Shpinics+ algorithm and renamed as SLH-DSA, i.e., StateLess Hash-based Digital Signature Algorithm). The fourth standard, FIPS 206 (based on FALCON, dubbed as FN-DSA - FFT over NTRU-Lattice-Based Digital Signature Algorithm), is under draft consideration, and the fifth standard, FIPS 207 (HQC-KEM, i.e., Hamming Quasi-Cyclic Key Encapsulation Mechanism), is under development at the time of the writing of this paper. NIST suggests starting the migration without waiting in a statement \textit{``With the release of the first three final PQC standards, organizations should begin migrating their systems to quantum-resistant cryptography''}  \cite{nistpqcproject}. 

With this in mind, large content delivery networks (CDNs) and cloud service providers such as Cloudflare, AWS, Microsoft, Google, and Akamai, etc. are already adopting PQC and migrating their existing encryption mechanisms to post-quantum algorithms. In March 2024, as reported by Cloudflare \cite{westerbaan2024pqc},  only \textit{``around two percent of all TLS 1.3 connections established with Cloudflare are secured with post-quantum cryptography''}; but in October 2025, Cloudflare claimed in a report \cite{westerbaan2025pqc} that \textit{``over 50\% (and rising!) of human traffic is protected against store-now/decrypt-later, making post-quantum key agreement the new security baseline for the Web''}. Also, Google has announced a 2029 timeline to secure the quantum era with post-quantum cryptography migration \cite{adkins2026pqc}. Thus, it is high time to conduct a real-world data-driven empirical analysis of PQC readiness across TLS and SSH by assessing the PQC deployment status. 

For this work, we selected the Secure Shell (SSH) protocol for PQC adoption assessment given its importance in cybersecurity and its consideration as a significant vector for privilege escalation, where a low-privileged attacker gains root access and can perform lateral movements within a network. Looking into the state-of-the-art of PQC, several studies highlight the need for PQC readiness in enterprise networks \cite{powell2026quantum, qureshi6045163risk, kundu2025quantum}. In a recent study, Mallick et al. \cite{mallick2026studypostquantumstatus} surveyed nine widely deployed protocols for PQC adoption feasibility - TLS, IPsec, BGP, DNSSEC, SSH, QUIC, OpenID Connect, OpenVPN, and the Signal Protocol. For each protocol, they provided cryptographic reliance, post-quantum migration efforts to date, and the challenges that still remain. 

Sikeridis et al. \cite{sikeridis2020post} provide a performance evaluation of the NIST signature algorithm candidates and investigate the imposed latency and throughput on TLS 1.3 connections, along with analyzing the trade-off between lengthy Post-Quantum (PQ) signatures and computationally heavy PQ cryptographic operations. Demir et al. \cite{demir2025performance} provide a performance study of CRYSTALS-Kyber and CRYSTALS-Dilithium by benchmarking execution times across cryptographic operations such as key generation, encapsulation, decapsulation, signing, and verification, and assessing hardware acceleration benefits. Hanna et al. \cite{hanna2025comprehensive} analyzed the viability of Post-Quantum algorithms in TLS to power-constrained IoT devices, focusing on two widely used IoT network protocol stacks (Bluetooth Low Energy (BLE) and Wi-Fi). 

These works focus on PQC deployment challenges and performance evaluation of the available algorithms in terms of latency, throughput, time to deploy, and challenges to face during migration, but there is very limited research on the status of real-world PQC deployment, specifically on SSH servers. In fact, to the best of our knowledge, this paper is the first of its kind to explore real-world large-scale data-driven analysis of PQC adoption assessment on SSH servers. Within the scope of this report, we have looked into several research aspects: 1) The full span of observed PQC algorithms in SSH servers, 2) PQC adoption status at the individual server level as well as entity-level, and 3) PQC adoption variations by organizations' industry, revenue, employee size, and digital size (such as IPv4 count or SSL active count). To note here, we use the term `server' or `host' to represent a unique SSH-enabled IP address, and the term `entity' represents an organization that hosts the IP address. A single entity may host multiple SSH servers. Several notable contributions of this work are as follows:
\begin{enumerate}[label=\roman*)]
  \item Discussion on PQC algorithms fingerprinting process in SSH servers, 
  \item Detailed host-level PQC adoption assessment, and 
  \item Comprehensive insights on organization-wide PQC adoption status based on several metrics such as digital and financial size metrics, as well as industry sector and geographic diversification.
\end{enumerate}

We organize the rest of the paper as follows. Section \ref{sec:2} discusses the data collection and processing methods of the PQC key exchange algorithm for the SSH servers. Section \ref{sec:3} is the host-level PQC adoption analysis, followed by the entity-level PQC adoption analysis in Section \ref{sec:4}. Section \ref{sec:5} provides the key findings. Finally, Section \ref{sec:6} concludes the paper.


\section{Data Collection \& Processing}\label{sec:2}
In this section, we present the data collection and processing methodologies of key exchange algorithms pertinent to the SSH protocol. There are several ways we can collect the KEX algorithms. We provide the details below. 



Key exchange (KEX) (aka `key encapsulation') algorithms are cryptographic methods that allow two parties to establish a shared secret key over an insecure channel. We can use the SSH client (e.g., OpenSSH) to collect the list of KEX algorithms for a specific target server on port 22 (SSH), but this method is not suitable for a large-scale scan of PQC key exchange algorithms. Large-scale, industry-wide data collection is challenging due to resource limitations and security restrictions. Bitsight has built an in-house internet scanning platform called GROMA that maintains a continuous global survey of the public-facing Internet. More details are available here \cite{bitsight2024groma}. 


\begin{figure}[!b]
{
\centering

\begin{lstlisting}[
  language=json,
  %escapeinside={(*}{*)} % Defines (* and *) as escape delimiters,
  basicstyle=\footnotesize\ttfamily, % Smaller font to "squeeze"
      lineskip=-1.5pt,                   % Tightens vertical space between lines
      frame=single,
      xleftmargin=0em,                   % Pulls the frame in
      xrightmargin=0em,
      aboveskip=0pt,                     % Removes space above the listing inside figure
      belowskip=0pt,                     % Removes space below the listing inside figure
      escapeinside={(*}{*)}
]
{
  "ip": "(*\color{black}\rule{2.5cm}{1.5ex}*)",
  "data": {
    "ssh": {
      "status": "success",
      "protocol": "ssh",
      "port": 22,
      "result": {
        "server_id": {
          "raw": "SSH-2.0-OpenSSH_10.2",
          "version": "2.0",
          "software": "OpenSSH_10.2"
        },
        "server_key_exchange": {
          "cookie": "(*\rule{4cm}{1.5ex}*)",
          "kex_algorithms": [
            "mlkem768x25519-sha256",
            "sntrup761x25519-sha512",
            "sntrup761x25519-sha512@openssh.com",
            "curve25519-sha256",
            "curve25519-sha256@libssh.org",
            "ecdh-sha2-nistp256",
            "ecdh-sha2-nistp384",
            "ecdh-sha2-nistp521",
            "ext-info-s",
            "kex-strict-s-v00@openssh.com"
          ],
          "host_key_algorithms": [
            "rsa-sha2-512",
            "rsa-sha2-256",
            "ecdsa-sha2-nistp256",
            "ssh-ed25519"
          ],
          "client_to_server_ciphers": [
            "aes256-gcm@openssh.com",
            "aes256-ctr",
            "aes128-gcm@openssh.com",
            "aes128-ctr"
          ],
...
}

\end{lstlisting}
\caption{Example of key exchange algorithm extraction using an open-source tool.}\label{fig:1}
}
\end{figure}




For demonstration purposes, we present a partial output for an SSH scan on a specific target (anonymized) using an open-source tool in Figure \ref{fig:1}. Here the \textbf{server\_key\_exchange} field contains the \textbf{kex\_algorithms} supported by the server. For our analysis, we use GROMA data to extract the key exchange algorithms on port 22. Later, we map the scanned IP addresses to entities to perform entity-level analyses. The entity mapping process maps the entity IDs corresponding to the IP addresses or domains, where each entity is a real-world organization. The mapping process relies on the use of an AI-powered asset mapping and attribution engine called `GIA'. More details on the entity mapping process and GIA are available here \cite{bitsight2024gia}. We collect firmographic data (i.e., revenue and employee count) from our service providers. We also leverage the digital metrics data (e.g., IPv4 count and SSL  count) from different internal data pipelines. 

As GROMA contains lots of different ports' scan data, we basically filtered the data on port 22 (service `ssh') for a full scan day. To extract the PQC key exchange algorithm variants, we created a custom RegEx expression to cover all the PQC algorithms mentioned in the Open Quantum Safe project \cite{oqs_liboqs_algorithms}.

Next, we discuss the main analyses on the assessment of PQC adoption status for the SSH protocol. We divide the discussion into two broad sections: server-level analysis \ref{sec:3} and entity-level analysis \ref{sec:4}. 

\section{Server-Level PQC Adoption Analysis}\label{sec:3}

This section provides details of the server-level PQC adoption status based on the data processed earlier. We divide the analyses into several subsections: 1) PQC KEX proportions, and 2) PQC only KEX algorithm variants.

\subsection{Server-Level PQC KEX Proportions} \label{subsec:3A}
We have a total of 1,548,074 records of SSH service in GROMA data for the data collection date (February 16, 2026). After data cleaning, we have a total of 1,523,998 unique records with both non-null unique IP addresses and the \textbf{`kex\_algorithms'} field. Based on the analysis, there are 688,814 records (server IP addresses) that have either one or multiple PQC algorithms listed in their \textbf{`kex\_algorithms'} field. Table \ref{table1} presents the proportions of PQC and non-PQC KEX algorithms. The proportion of PQC presence on these SSH servers is roughly 45.2\%. There are 835,184 records (54.8\%) (i.e., IP addresses) that do not have any PQC algorithms in the \textbf{`kex\_algorithms'} field (i.e., they support only classical algorithms). Thus, the share of the PQC algorithm adoption is significant, as we see in Table \ref{table1}. To mention the \textbf{`kex\_algorithms'} field is an array of supported algorithms, which may include multiple PQC and non-PQC algorithms at the same time. For example, as shown in Figure \ref{fig:1}, we have three different PQC variants, and the server is counted as PQC enabled (i.e., `true’) if any of the PQC algorithm variants is present in that array of kex algorithms as defined by the RegEx earlier complying with list of key encapsulation algorithms provide by the ``Open Quantum Safe Project''  \cite{oqs_liboqs_algorithms}. 

\begin{table}[!b]
\vspace{-4pt}
    \centering
    \caption{Proportions of Server-Level Key Exchange Algorithms}
    \label{table1}
    \small
    \begin{tabularx}{\columnwidth}{Xcr} 
        \toprule
        \textbf{KEX Algorithm} & \textbf{Count} & \textbf{Percentage (\%)} \\ 
        \midrule
        PQC                    & 688,814        & 45.2 \\ 
        Non-PQC                & 835,184        & 54.8 \\ 
        \bottomrule
    \end{tabularx}
\end{table}

\subsection{Server-Level PQC Only KEX Algorithm Variants} \label{subsec:3B}

\begin{table}[!t]
    \centering
    \caption{Observed Server-Level PQC KEX Algorithm Variants}
    \label{table2}
    \scriptsize 
    \addtolength{\tabcolsep}{-3pt} 
    \begin{tabularx}{\columnwidth}{l rr} 
        \toprule
        \textbf{PQC KEX Variants} & \textbf{Count} & \textbf{Percentage (\%)} \\ 
        \midrule
        \texttt{sntrup761x25519-sha512@openssh.com} & 681,490 & 44.86 \\ 
        \texttt{sntrup761x25519-sha512}             & 168,414 & 11.08 \\ 
        \texttt{mlkem768x25519-sha256}             & 88088 & 5.79 \\ 
        \texttt{sntrup4591761x25519-sha512@tinyssh.org} & 1,035 & 0.07 \\ 
        \texttt{mlkem1024nistp384-sha384}             & 680 & 0.04 \\ 
        \texttt{mlkem768nistp256-sha256}             & 358 & 0.02 \\ 
        \texttt{mlkem768brainpoolp256r1-sha256}             & 11 & 0.00 \\ 
        \texttt{curve25519-frodokem1344-sha512@ssh.com}     & 2 & 0.00 \\ 
        \texttt{ecdh-nistp521-kyber1024-sha512@ssh.com}     & 2 & 0.00 \\ 
        \bottomrule
    \end{tabularx}
\end{table}

Table \ref{table2} presents the percentages of each PQC variant in the SSH data as we observed. From the illustration, it seems that the hybrid streamlined NTRU prime (sntrup761) with X25519, along with SHA-512, contains major proportions of the observed PQC variants. We also observed different variants of ML-KEM in the KEX algorithms, e.g., mlkem768x25519-sha256, mlkem1024nistp384-sha384, mlkem768nistp256-sha256, and mlkem768brainpoolp256r1-sha256. Thus, the SNTRUP seems to be the dominant PQC algorithm, followed by the ML-KEM variants. The \textbf{`kex\_algorithm'} field contains multiple supported algorithms in an array; these percentages are not mutually exclusive. Meaning both SNTRUP or ML-KEM encryption algorithms variants may be present in the same server that supports SSH on port 22, based on the server-side PQC deployment. 

There could be some confusion regarding variants of sntrup761x25519 as found in Table \ref{table2}. While sntrup761x25519-sha512 and sntrup761x25519-sha512@openssh.com are basically the same algorithm, they were renamed slightly differently because of the standardization practice. When OpenSSH 9.0 first introduced the post-quantum hybrid in 2022, the algorithm hadn't been officially standardized by the Internet Engineering Task Force (IETF) then. To avoid clashing with future standards, OpenSSH used its own private namespace by adding `@openssh.com'. As per our knowledge, this is a standard practice that any algorithm name ending in a domain name is a private extension specific to that vendor. As the algorithm became widely adopted and moved toward official standardization, the name sntrup761x25519-sha512 was established. As a reference regarding different naming conventions of SNTRUP761 variants, readers are encouraged to review the acknowledgement sections in this IETF draft \cite{josefsson-ntruprime-ssh-02}.

\section{Entity-Level PQC Adoption Analysis}\label{sec:4}
As illustrated earlier, we perform an entity-mapping technique to map the IP addresses of the SSH servers to specific entities and perform a deduplication process to reduce any duplicate count. Next, we present different analyses on entity-level PQC adoption.

\subsection{Entity-Level PQC Proportions} \label{subsec:4A}
In the SSH data collected for the selected scan day, we have 27,012 unique entities; among them, 10788 entities support PQC in 1 or more of their SSH servers (IP addresses), and 16,224 entities do not support any of the PQC algorithm variants on any of their IPs on port 22. Thus, the percentage of entities with PQC adoption in 1 or more IP addresses is 39.9\%, and entities without any PQC adoption on any of their servers is 60.1\%. Thus, roughly around 40\% of the entities support PQC on at least 1 IP address (server) associated with them.

\begin{table}[!t]
    \centering
    \caption{Proportions of Entity-Level Key Exchange Algorithms}
    \label{table3}
    \small
    \begin{tabularx}{\columnwidth}{Xcr} 
        \toprule
        \textbf{KEX Algorithm} & \textbf{Count} & \textbf{Percentage (\%)} \\ 
        \midrule
        PQC                    & 10,788        & 39.9 \\ 
        Non-PQC                & 16,224        & 60.1 \\ 
        \bottomrule
    \end{tabularx}
\end{table}

\begin{figure}[!b]
{
\centering

\includegraphics[width=0.48\textwidth, trim=0 0 0in 0.25in, clip, keepaspectratio]{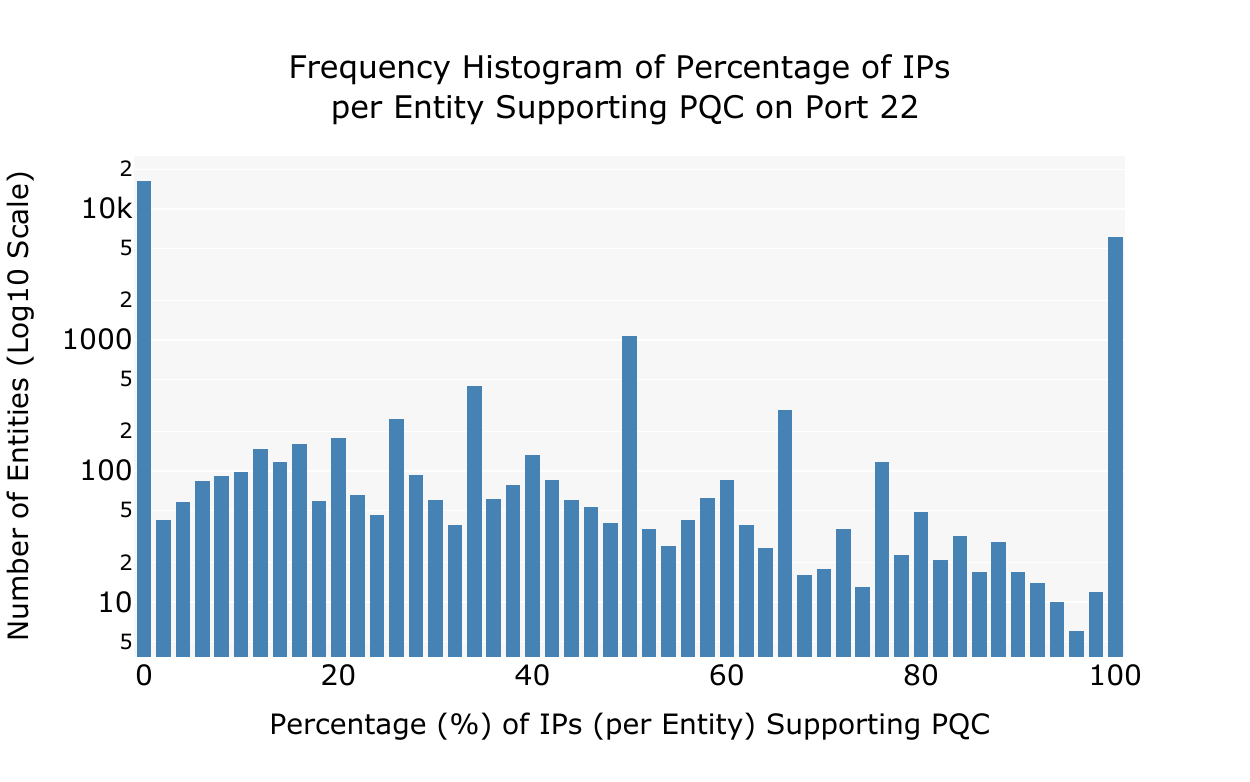}
\caption{Frequency histogram of percentage of IPs supporting PQC per entity.}\label{fig:3}
}
\end{figure}

As entities may have completely different numbers of IPs that belong to them, the next research question that we explored is \textit{``what percentage of IPs per entity support PQC on port 22?''}. Figure \ref{fig:3} presents an analysis of the percentage of IPs per entity supporting PQC. This histogram provides a high-level view of how PQC adoption is distributed across different entities (i.e., organizations). The X-axis represents the percentages of IPs supporting PQC, and the Y-axis represents the number of entities on a log10 scale. 

From this analysis, we find around 60\% of entities (16,236 out of 27,012) fall into the bin of $<$1\% of IPs supporting PQC, thus having the 'legacy group' majority. This indicates that for the vast majority of organizations, PQC for SSH is not yet a deliberate configuration across their fleet, consisting mostly of ad-hoc or testing instances. There is also a full-migration group, which is around 23\% (6,127 out of 27,012) of entities that show near 100\% PQC support, suggesting a successful, systemic rollout across their entire SSH infrastructure. While a middle-ground exists (entities with 1$\sim$20\% of IPs supporting PQC), the log10 scale highlights a massive disparity; the legacy group is $\sim$150x larger than those in the active transition bins ($\sim$100 entities per bin). This shows a massive adoption gap, which also explains that most organizations remain entirely reliant on classical cryptography on port 22, while a smaller, proactive segment has already fully transitioned. 

\begin{figure*}[!t] 
    \centering
    \includegraphics[width=\textwidth, trim=0.5in 0 0.25in 0, clip]{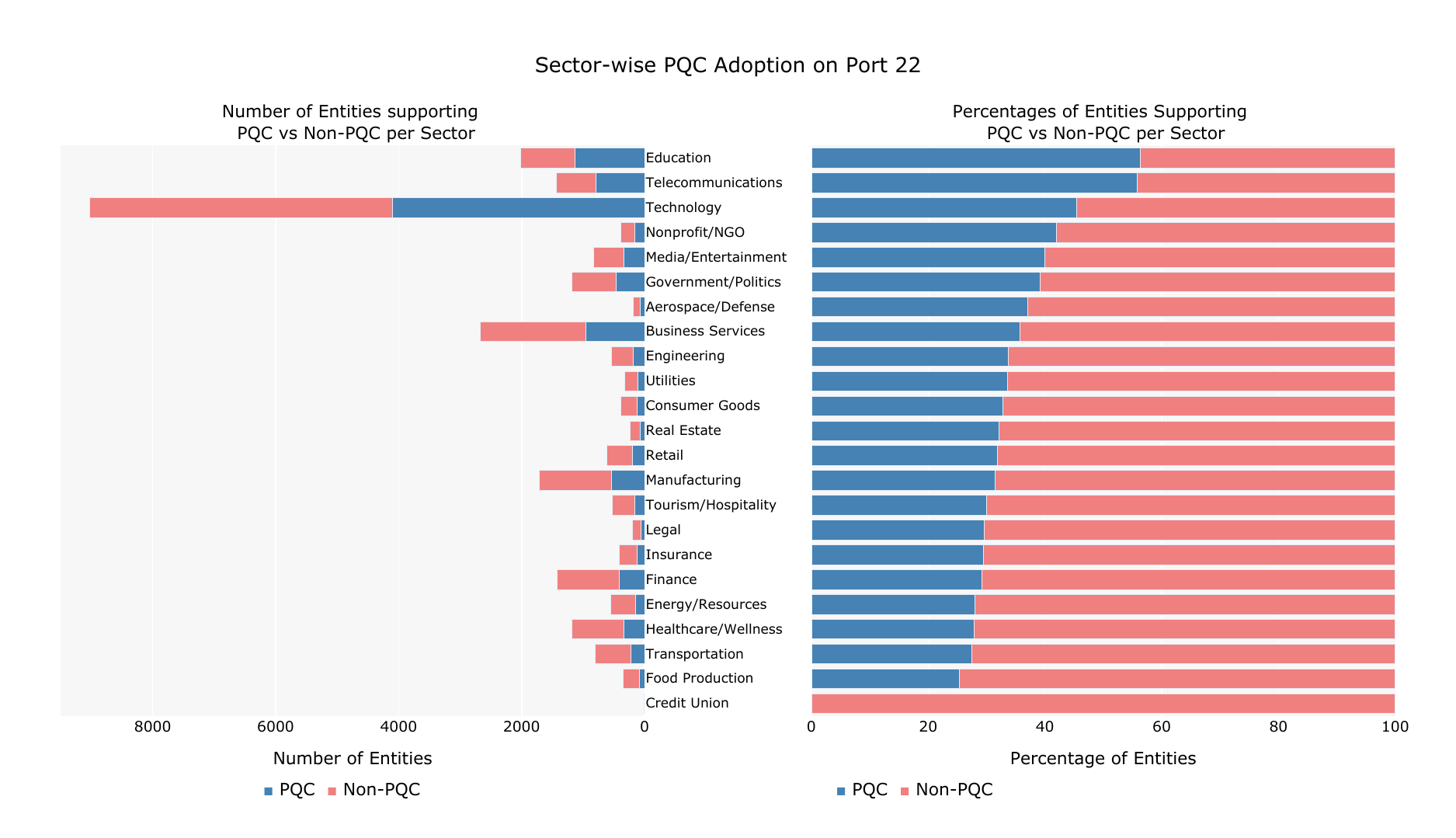}
    \caption{Entities sector-wise PQC adoption on port 22.}
    \label{fig:4}
\end{figure*}

\subsection{Sector-Wise PQC Adoption} \label{subsec:4B}

In the data, we have a total of 23 sectors associated with the entities. Figure \ref{fig:4} illustrates the PQC adoption in different sectors in a mirrored-style visualization where it contrasts the absolute volume of PQC adoption by the entities per sector on the left with the relative maturity (\% of entities) per sector on the right. As we observed, the `Technology' sector has one of the largest absolute counts of PQC-supporting entities. The sectors ‘Technology’, ‘Business Services’, ‘Manufacturing’, ‘Education’ and ‘Finance’ represent the bulk of the data points, indicating where the most infrastructure is concentrated.

The right part of the figure shows the percentages of entities supporting PQC within each sector. `Education' and `Telecommunications' are the frontrunners, with over 50\% of entities supporting PQC. This suggests that these sectors are more proactive in updating their SSH configurations. Most industrial and professional sectors (e,g., Finance, Healthcare, Legal) hover between 25\% and 35\% adoption, representing a standard `early adoption' phase across the enterprise world.

While `Technology' has the most PQC-ready entities by count (on the left), it ranks 3rd in percentage (on the right). This suggests that while tech companies are leading the migration, the sheer size of their legacy infrastructure possibly makes it difficult to roll out a 100\% PQC on SSH servers. Sectors like `Education' often adopt new standards faster due to research initiatives and modern greenfield deployments, whereas sectors like `Finance' or `Utilities' often have slower change management cycles due to strict regulatory requirements. The transition from the top (`Education' or `Telecom' industry) to the bottom (`Credit Unions') shows a sharp decline. The bottom half of the list is remarkably consistent in percentages of entities with PQC adoption, with most sectors stagnant at roughly one-third adoption. This illustrates the existence of a sector-wide adoption gap. 

\subsection{Entity Size-Based PQC Adoption} \label{subsec:4C}

\begin{figure*}[!t] 
    \centering
    \includegraphics[width=0.85\textwidth, trim=0in 0 0.15in 0, clip]{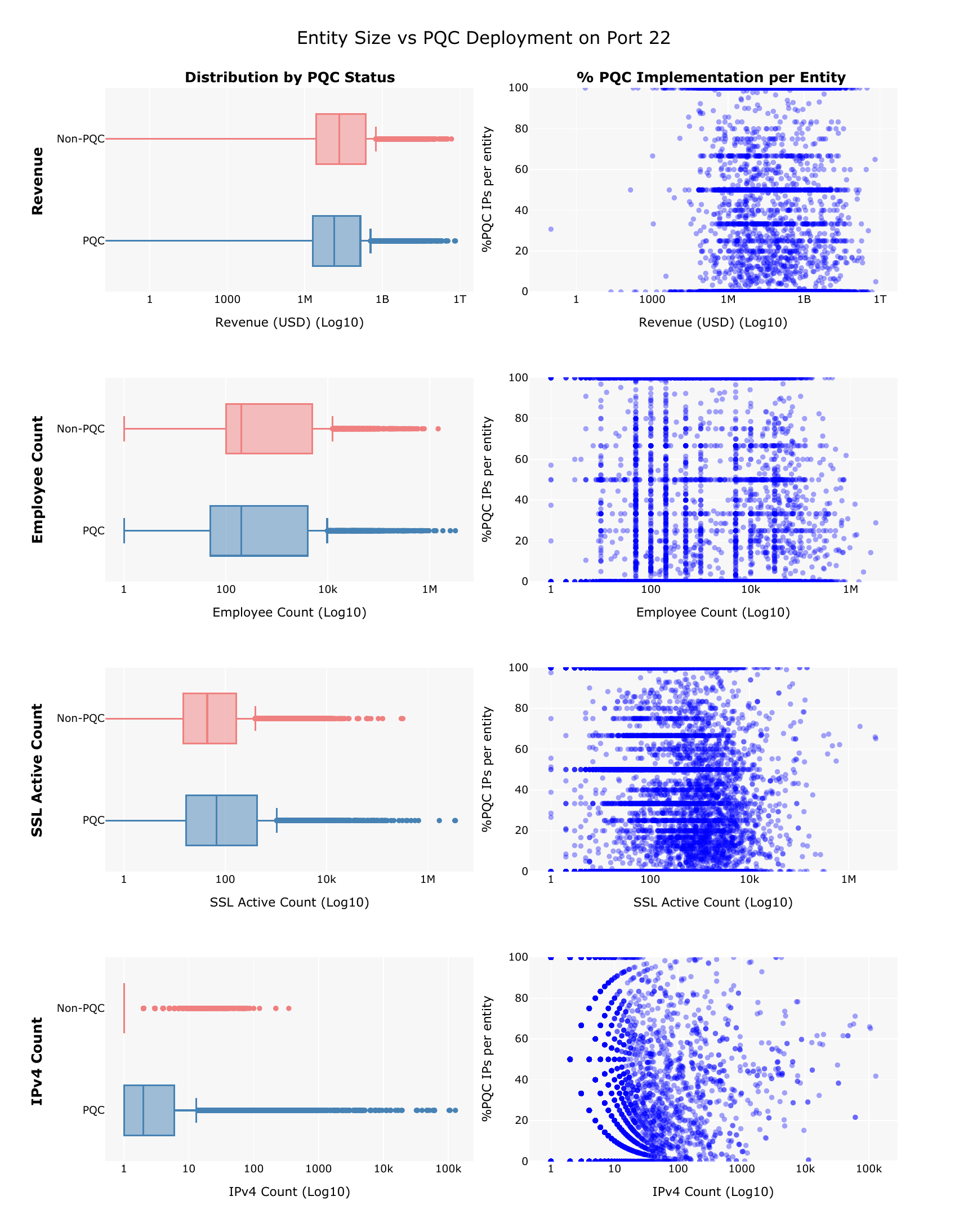}
    \caption{Entities financial and digital size metrics-based PQC adoption in SSH.}
    \label{fig:5}
\end{figure*}

We perform entity size-based analysis to understand the relationship between entity size metrics and PQC adoption on SSH. We selected two broad classes of entity size metrics: financial metrics and digital size metrics. The financial metrics include two sub-metrics: the organization's revenue (in USD) and the number of employees. The digital size metric includes `SSL\_ACTIVE\_Count' and IPv4 address count. To give some context, \textbf{`SSL\_ACTIVE\_Count'} is basically the number of unique SSL evidence keys (i.e., unique findings) graded over the past 60 days that are believed to be still online. An evidence key is either an `IP:PORT' or `Hostname:Port' information. 

We present the analysis in a 4x2 grid view of the selected entity size metrics in Figure \ref{fig:5} where each row represents a metric, either financial or digital size of the entities. The row metrics are revenue in USD, employee count, SSL active count, and IPv4 count, respectively. The first column of each row indicates the Boxplot distribution of PQC and non-PQC entities over those size metrics. The second column represents the percentage of IPs per entity with PQC adoption vs the entity size metrics. Thus, the first column compares the distribution of PQC-supported entities with that of non-PQC-supported entities. And the second column shows the adoption depth of PQC entities with the percentage of IPs per entity supporting PQC. The X-axes are in Log10 scale because of the large range of those metrics. From Figure \ref{fig:5}, we can summarize the major findings on this analysis as follows:

\begin{figure*}[!t] 
    \centering
    \includegraphics[width=0.9\textwidth, trim=0.5in 1in 0.25in 0, clip]{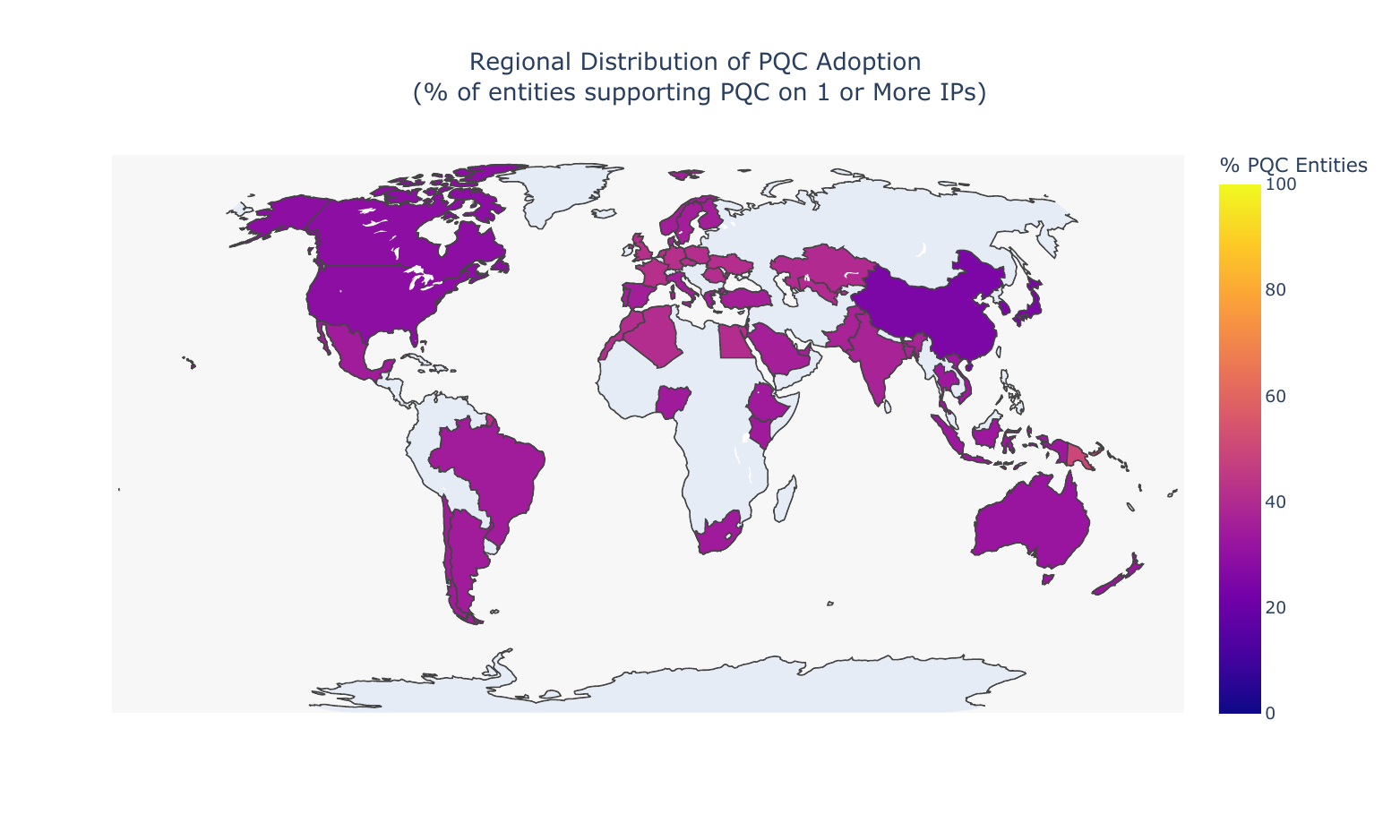}
    \caption{Regional distribution of PQC adoption in SSH.}
    \label{fig:6}
\end{figure*}

\begin{itemize}[leftmargin=*, labelsep=1em]
    \item \textbf{Decoupling of revenue and PQC adoption:} PQC adoption in SSH infrastructure is not restricted to a specific revenue bracket or employee count. Both small-scale organizations and multi-billion dollar enterprises show similar median distributions, indicating that revenue (budget) and headcount are not the primary barriers to adopt post-quantum security.

    \item \textbf{Digital footprint as a driver:} PQC adoption is widespread across organizational sizes when measured by digital size metrics (SSL active and IPv4 counts). However, entities with more mature web presences (i.e., higher SSL active counts) show a slightly higher propensity to initiate PQC deployment.

    \item \textbf{Divergence in PQC implementation depth:} There is a clear divergence in PQC implementation depth:
    \begin{itemize}[label=$\circ$] 
        \item \textit{Small Entities:} Benefit from digital agility, making it easier to achieve 100\% organization-wide PQC coverage.
        \item \textit{Large Enterprises:} While they are active adopters, they face a complexity penalty. As the number of IPv4 addresses increases, the density of entities achieving 100\% PQC support thins out, suggesting significant challenges in migrating legacy SSH infrastructure to PQC at scale.
    \end{itemize}
\end{itemize}

Next, we discuss the geographic diversification of PQC adoption in Section \ref{subsec:4D}. 

\subsection{Geographic Distribution of PQC Adoption} \label{subsec:4D}
There are a total of 16 regions in the processed data. We have a total of 15,820 entities that have a region value; among them, 9,709 entities have no IPs supporting PQC, and 6,111 entities have one or more PQC-enabled IPs. A broad regional distribution of relative adoption density (percentages of entities supporting PQC on 1 or more IPs on port 22) is given in Figure \ref{fig:6} with approximate geographic mapping. 

The color scale highlights a clear concentration of PQC-ready entities in specific regions. Regions like North America, Western Europe, and parts of Northern Africa/Middle East (e.g., Morocco, Egypt, Saudi Arabia) show adoption rates in the 30\% to 50\% range. This suggests that in these regions, nearly one out of every two or three entities has begun their PQC transition. A vast majority of the active world, including Brazil, China, Australia, and India, falls into the 20\% to 35\% bracket. This indicates a baseline global adoption level where PQC is present but not yet the majority standard. This also reflects the standard `technology lag' where newer cryptographic standards are often first deployed in high-density tech hubs before trickling down to more emerging digital economies. These findings demonstrate that while PQC has a foothold globally, we are still far from a `Quantum-Secure World' by default on SSH, as no major region has yet crossed the 80$\sim$100\% (yellow colored) threshold.

\section{Key Findings}\label{sec:5}
We summarize the key findings of this analysis below:
\begin{itemize}[leftmargin=2.5em, labelsep=1em]
    \item \textbf{PQC adoption status:} Overall, while host-level support is nearing a majority ($\sim$45\%), entity-level adoption lags, indicating that many organizations have yet to initiate their post-quantum transition on SSH.
    
    \item \textbf{Dominant PQC variants:} The continued dominance of sntrup761 over NIST-standardized ML-KEM stems from early OpenSSH-based deployments that began long before NIST finalized the PQC encryption standards in 2024. This also implies that most organizations aren't adopting PQC intentionally. They are possibly unintentionally PQC-ready simply because, at some point, they upgraded their SSH server (likely OpenSSH) to a version that supports sntrup761 by default.

    \item \textbf{Slow PQC migration:} The polarized distribution suggests a ``binary'' adoption pattern where entities either fully commit to infrastructure-wide PQC migration on SSH or remain entirely tethered to legacy classical cryptographic systems. Others are implementing PQC on SSH gradually, but the pace is slow, as observed from the histogram of the percentages of IPs supporting PQC per entity. 
    
    \item \textbf{Sector-wide PQC adoption:} Adoption maturity is currently driven by research-oriented and infrastructure-heavy sectors like `Education' and `Telecommunications', while highly regulated industries like `Finance', `Healthcare' and `Legal' still remain in the experimental phase.

    \item \textbf{Digital footprint drives PQC adoption:} Analysis shows that digital maturity, not financial scale, is the primary driver of PQC adoption; revenue and employee counts show nearly identical distributions for adopters and non-adopters alike. While higher SSL and IPv4 counts correlate with PQC support, adoption density thins as network size increases, reflecting the complexity of migrating legacy infrastructure. Ultimately, PQC readiness is a function of digital footprint and network agility rather than financial capacity.
    
    \item \textbf{Gap in global-level PQC adoption:} The stark geographic disparity underscores that PQC readiness is currently concentrated within regions possessing advanced digital infrastructure and robust cybersecurity regulatory frameworks.
    
\end{itemize}

\section{Conclusions}\label{sec:6}

We perform a comprehensive empirical analysis of the current PQC adoption status on SSH infrastructure based on large-scale industry-wide real-world data. The analysis shows some promising insights into the PQC deployment status. We explored the research questions around PQC key exchange algorithm variants and their observed share in SSH servers. We also analyzed other dimensions of PQC-related analysis, such as industry sectors, geographic regions, financial and digital metrics, and distribution of PQC-enabled entities. The analysis confirms that PQC has moved beyond theoretical discussion into a significant phase of real-world implementation within the SSH protocol. 

Our analysis of 1.52 million SSH-enabled hosts and 27,012 entities reveals that while the PQC transition is well underway, with 45.2\% of hosts and 39.9\% of entities supporting quantum-safe algorithms, the landscape remains deeply polarized. PQC adoption in SSH is characterized by a bimodal distribution: a proactive 23\% vanguard has achieved near-total migration, while a 60\% legacy majority still maintains negligible PQC support. Market consolidation is evident, with SNTRUP variants (notably OpenSSH’s implementation) dominating at 44.9\%, followed by ML-KEM at 5.8\%. Crucially, the study finds that digital footprint (IPv4 count and SSL counts) is a better predictor of PQC adoption than financial scale. Large-scale networks also face a significant last-mile complexity challenge. Geographically, while North America and Western Europe lead the adoption density, the lack of any region exceeding the 80\% threshold underscores that despite a strong global foothold, a `secure-by-default' PQC state on SSH infrastructures has yet to be realized.

\section*{Disclaimer}
This work leverages AI tools to help with some of the data processing tasks. The paper is solely written by the authors without the help of AI tools. The data used for this research are proprietary to Bitsight. 
 
\bibliographystyle{IEEEtran}
\bibliography{reference}





\end{document}